\documentclass[12pt,draftcls,onecolumn]{IEEEtran}

\makeatletter
\def\ps@headings{%
\def\@oddhead{\mbox{}\scriptsize\rightmark \hfil \thepage}%
\def\@evenhead{\scriptsize\thepage \hfil\leftmark\mbox{}}%
\def\@oddfoot{}%
\def\@evenfoot{}}
\makeatother 

\usepackage{amsfonts}
\usepackage{multirow}
\usepackage[tbtags]{amsmath}
\usepackage{graphicx,subfigure}
\usepackage{color}      
\usepackage{epsfig}
\usepackage{amssymb}
\usepackage{tipa}
\usepackage{bbm}

\long\def\comment#1{}

\newcommand{\beq}{\begin{equation}}
\newcommand{\eeq}{\end{equation}}
\newcommand{\beqno}{\begin{equation*}}
\newcommand{\eeqno}{\end{equation*}}

\newcommand{\bes}{\begin{split}}
\newcommand{\ees}{\end{split}}
\newcommand{\bdm}{\begin{displaymath}}
\newcommand{\edm}{\end{displaymath}}

\newtheorem{definition}{Definition}

\newcommand{\bd}{\begin{definition}}
\newcommand{\ed}{\end{definition}}

\newcommand{\bv}{\begin{vugraph}}
\newcommand{\ev}{\end{vugraph}}
\newcommand{\bi}{\begin{itemize}}
\newcommand{\ei}{\end{itemize}}
\newcommand{\ben}{\begin{enumerate}}
\newcommand{\een}{\end{enumerate}}

\newcommand{\bean}{\begin{eqnarray*} }
\newcommand{\eean}{\end{eqnarray*} }
\newcommand{\bea}{\begin{eqnarray} }
\newcommand{\eea}{\end{eqnarray} }
\newcommand{\nn}{\nonumber}
\newcommand{\ba}{\begin{array} }
\newcommand{\ea}{\end{array} }

\begin{document}

\title{Maximum-Likelihood Sequence Detector  for
Dynamic Mode High Density Probe Storage }


\author{Naveen Kumar, Pranav Agarwal, Aditya Ramamoorthy and Murti V. Salapaka \thanks{Naveen Kumar and
Aditya Ramamoorthy are with Dept. of Electrical and Computer Engg.
at Iowa State University, Ames IA 50011 (email: \{nk3,
adityar\}@iastate.edu). Pranav Agarwal and Murti V. Salapaka are
with the Dept. of Electrical and Computer Engg. at University of
Minnesota, Minneapolis, MN 55455 (email: \{agar0108,
murtis\}@umn.edu). The material in this work has appeared in part at
IEEE GlobeCom 2009 and in part at CISS 2008.}} \maketitle
\vspace{-2cm}
\begin{abstract}\vspace{-.2cm}

There is an increasing need for high density data storage devices
driven by the increased demand of consumer electronics. In this
work, we consider a data storage system that operates by encoding
information as topographic profiles on a polymer medium. A
cantilever probe with a sharp tip (few nm radius) is used to create
and sense the presence of topographic profiles, resulting in a
density of few Tb per in.$^2$. The prevalent mode of using the
cantilever probe is the static mode that is harsh on the probe and
the media. In this article, the high quality factor dynamic mode
operation, that is less harsh on the media and the probe, is
analyzed. The read operation is modeled as a communication channel
which incorporates system memory due to inter-symbol interference
and the cantilever state. We demonstrate an appropriate level of
abstraction of this complex nanoscale system that obviates the need
for an involved physical model. Next, a solution to the maximum
likelihood sequence detection problem based on the Viterbi algorithm
is devised. Experimental and simulation results demonstrate that the
performance of this detector is several orders of magnitude better
than the performance of other existing schemes.

\end{abstract}
\vspace{-.5cm}
\section{Introduction}
\label{sec:introduction} Present day high density storage devices
are primarily based on magnetic, optical and solid state
technologies.
Advanced signal processing and detection techniques have played an
important role in the design of all data storage
systems~\cite{woodp86,moonsp98,prml92,moon90,moon01,kavcic2000,forney1972}.
Indeed techniques such as partial-response max-likelihood
\cite{prml92,thaparmag87,woodp86} were responsible for significantly
improving magnetic disk technology.

In this work, we consider a promising high density storage
methodology which utilizes a sharp tip at the end of a micro
cantilever probe to create, remove and read indentations (see
\cite{vettiger02}). The presence/absence of an indentation
represents a bit of information. The main advantage of this method
is the significantly higher areal densities compared to conventional
technologies that are possible. Recently, experimentally achieved
tip radii near 5 nm on a micro-cantilever were used to create areal
densities close to 1
Tb/in$^2 ~\cite{vettiger02}.$  

A particular realization of a probe based storage device that uses
an array of cantilevers, along with the static mode operation is
provided in \cite{millipede03}. However, there are fundamental
drawbacks of this technique. In the static mode operation, the
cantilever is in contact with media throughout the read operation
which results in large vertical and lateral forces on the media and
the tip. 
Moreover, significant information content is present in the low
frequency region of the cantilever deflection and
it can be shown experimentally that the system gain at low frequency
is very small. Therefore, in order to overcome the measurement noise
at the output, the interaction force between the tip and the medium
has to be
large. This degrades the medium and the probe over time, resulting in reduced device lifetime.

The problem of tip and media wear can be partly addressed by using
the dynamic mode operation; particularly when a cantilever with a
high quality factor is employed. In the dynamic mode operation, the
cantilever is forced sinusoidally using a dither piezo. The
oscillating cantilever gently taps the medium and thus the lateral
forces are reduced which decreases the media
wear~\cite{lateralforcetapping}. Using cantilever probes that have
high quality factors leads to high resolution, since the effect of a
topographic change on the medium on the oscillating cantilever lasts
much longer (approximately $Q$ cantilever oscillation cycles, where
each cycle is $1/f_0$ seconds long and $Q$ and $f_0$ is the quality
factor and the resonant frequency of the cantilever respectively).
Moreover, the SNR improves as $\sqrt{Q}$~\cite{Wisendanger}.
However, this also results in severe inter-symbol-interference,
unless the topographic changes are spaced far apart. Spacing the
changes far apart is undesirable from the storage viewpoint as it
implies lower areal density. Another issue is that the cantilever
exhibits complicated nonlinear dynamics. For example, if there is a
sequence of hard hits on the media, then the next hit results in a
milder response, i.e., the cantilever itself has inherent memory,
that cannot be modeled as ISI. Conventional dynamic mode methods
described in~\cite{sahooSS05}, that utilize high-Q cantilevers are
not suitable for data storage applications. This is primarily
because they are unable to deal with ISI and the nonlinear channel
characteristics. The current techniques can be considered analogous
to peak detection techniques in magnetic storage \cite{moon90}.
In this work we demonstrate that these issues can be addressed by
modeling the dynamic mode operation as a communication system and
developing high performance detectors for it. Note that
corresponding activities have been undertaken in the past for
technologies such as magnetic and optical storage \cite{moonsp98},
e.g., in magnetic storage, PRML techniques, resulted in tremendous
improvements. In our work, the main issues are, (a) developing a
model for the cantilever dynamics that predicts essential
experimental features and remains tractable for data storage
purposes, and (b) designing high-performance detectors for this
model, that allow the usage of high quality cantilevers, without
sacrificing areal density. As discussed in the sequel, several
concepts such as Markovian modeling of the cantilever dynamics and
Viterbi detection in the presence of noise with
memory~\cite{kavcic2000}, play a key role in our approach.

\noindent \underline{ \it{Main Contributions:}} In this article, a
dynamic mode read operation is researched where the probe is
oscillated and the media information is modulated on
the cantilever probe's oscillations. 
It is demonstrated that an appropriate level of abstraction is
possible that obviates the need for an involved physical model. The
read operation is modeled as a communication channel which
incorporates the system memory due to inter-symbol interference and
the cantilever state that can be identified using training data.
Using the identified model, a solution to the maximum likelihood
sequence detection problem based on the Viterbi algorithm is
devised. Experimental and simulation results which corroborate the
analysis of the detector, demonstrate that the performance of this
detector is several orders of magnitude better than the performance
of other existing schemes and confirm performance gains that can
render the dynamic mode operation feasible for high density data
storage purposes.

Our work will motivate research for fabrication of prototypes that
are massively parallel and employ high quality cantilevers (such as
those used with the static mode~\cite{vettiger02} and intermittent
contact dynamic mode but
with low-Q~\cite{sahoo08}). 
In current prototypes, the cantilever detection is integrated into
the cantilever structure and the cantilevers are actuated
electrostatically. Even though the experimental setup reported in
this article uses  a particular scheme for measuring the cantilever
detection and for actuating the cantilever, the paradigm developed
for data detection is largely applicable in principle to other modes
of detection and actuation of the cantilever. The analysis criteria
primarily assume that high quality factor cantilevers are employed
and that a dynamic mode operation is pursued.

The article is organized as follows. In
Section~\ref{sec:physical_modeling}, background and related work of
the probe based data storage system is presented.
Section~\ref{sec:channel_modelling} deals with the problem of
designing and analyzing the data storage unit as a communication
system and finding efficient detectors for the channel model.
Section~\ref{sec:Simulation Results} and
Section~\ref{sec:Experimental Results} report results from
simulation and experiment respectively.
Section~\ref{sec:conclusions} provides the main findings of this
article and future work. \vspace{-.6cm}
\section{Background and related work.}\label{sec:physical_modeling}
\begin{figure}[t]
\centering
\begin{tabular}{cc}
\includegraphics[width=2.5in]{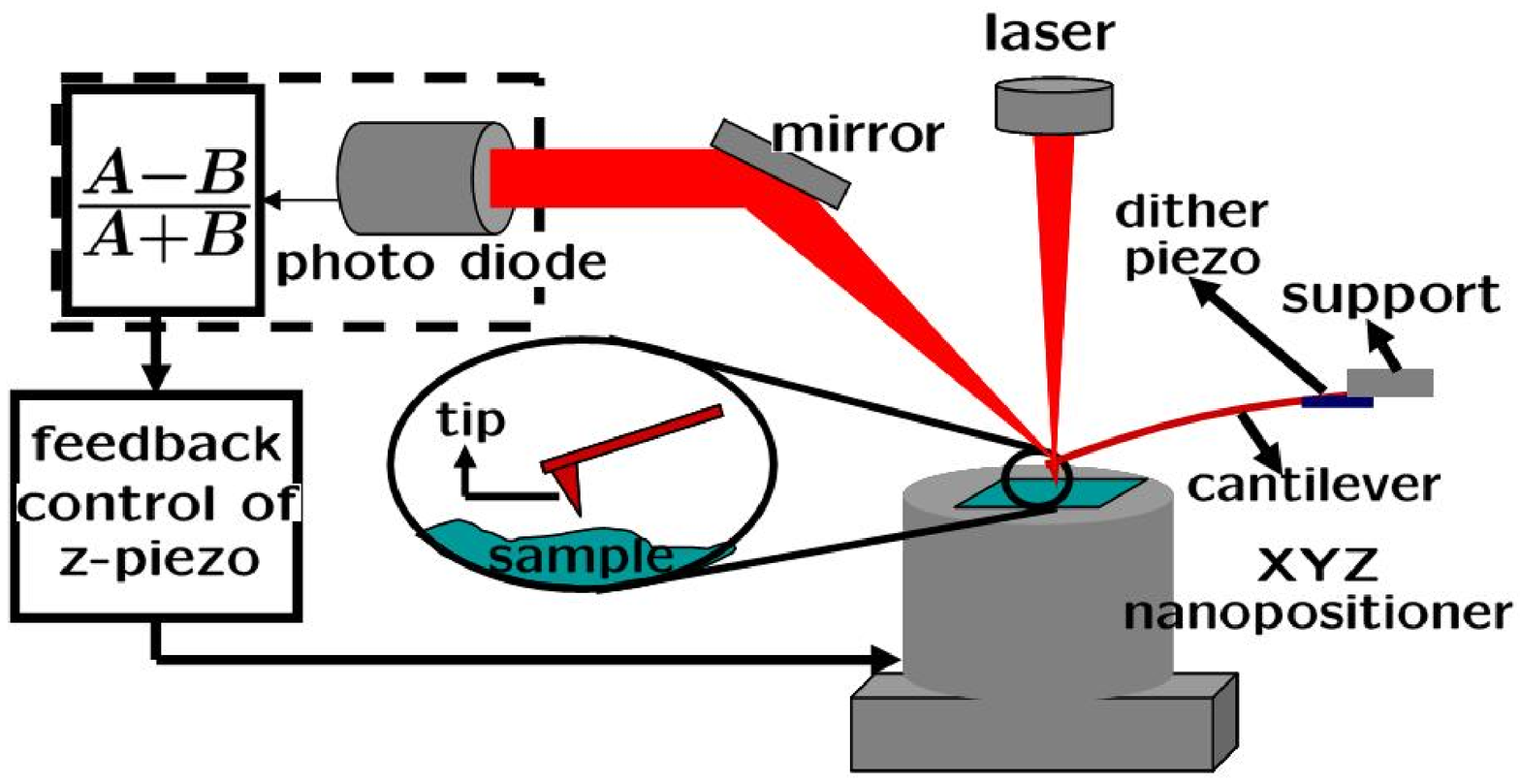} &\includegraphics[width=2.5in]{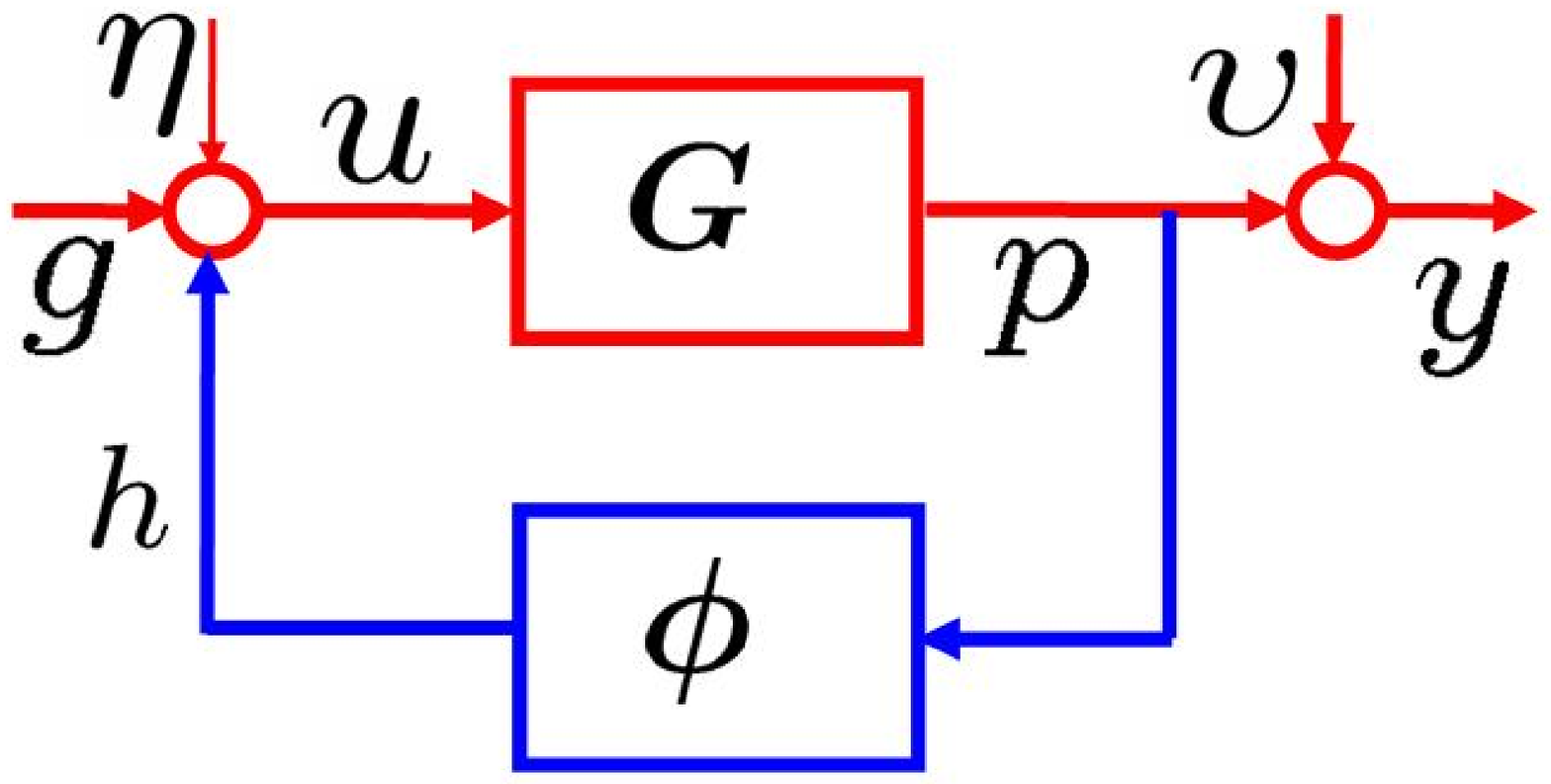}\\
(a) & (b)\\
\includegraphics[width=2.5in]{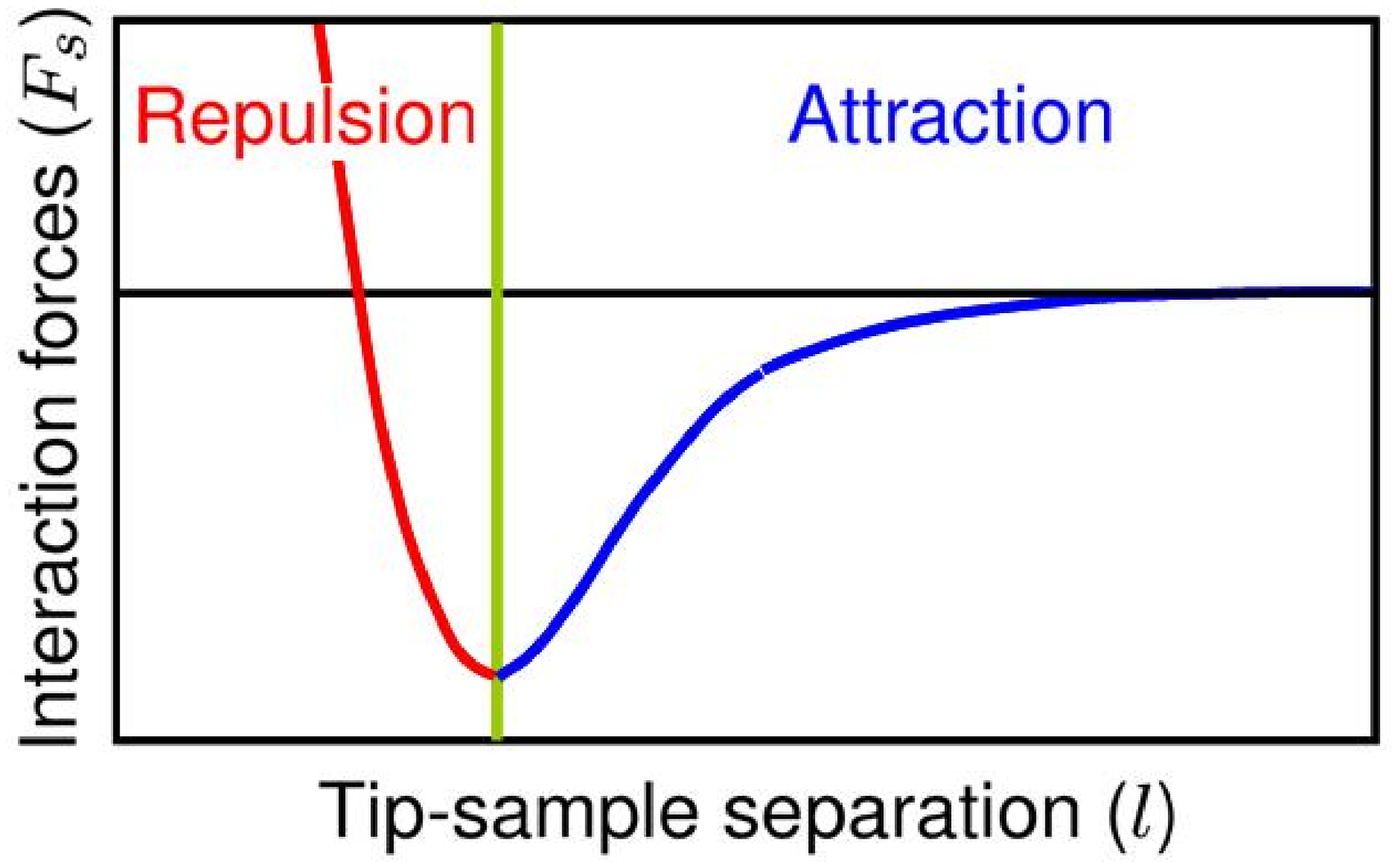}&\\(c) &
\end{tabular}
 \caption{\label{fig:afm}(a)  Shows the main components of a probe based storage device. The main probe is a cantilever with a tip at one end that interacts with the media. The support end can be forced using a dither piezo.
 The deflection of the tip-end is measured by a laser-mirror-photodiode arrangement. The controller
  employs the deflection measurement to keep the probe engaged with the media.
  (b)   Shows a block diagram representation of the
   cantilever system $G$ being forced by white noise ($\eta$),
    tip-media force $h$
    and the dither forcing $g$. The output of the block $G,$
    the deflection $p$ is corrupted
    by measurement noise $\upsilon$ that results in the measurement
    $y.$ Tip media force $h=\phi(p).$
     (c)
Shows the typical tip-media interaction forces of  weak long range
attractive forces and strong repulsive short range forces. }
\vspace{-.2cm}
\end{figure}

\vspace{-.2cm}
Probe based high density data storage devices employ a cantilever
beam that is supported at one end and has a sharp tip at another end
as a means to determine the topography of the media on which
information is stored. 
The information on the media is encoded in terms of topographic
profiles. A raised topographic profile is considered a high
bit and a lowered topographic profile is considered a low bit. 
There are various means of measuring the cantilever deflection. In
the standard atomic force microscope setup, which has formed the
basis of probe based data storage, the cantilever deflection is
measured by a beam-bounce method where a laser is incident on the
back of the cantilever surface and the laser is reflected from the
cantilever surface into a split photodiode. The photodiode collects
the incident laser energy
and provides a measure of the cantilever deflection (see
Figure~\ref{fig:afm}(a)). The advantage of the beam-bounce method is
the high resolution (low measurement noise) and high bandwidth (in
the 2-3 MHz) range. The disadvantage is that it cannot be easily
integrated into an operation where multiple cantilevers operate in
parallel. There are attractive measurement mechanisms that integrate
the cantilever motion sensing onto the cantilever itself. These
include piezo-resistive sensing \cite{chui98} and thermal sensing
\cite{durig05}. 
For the dynamic mode operation there are various schemes to actuate
the cantilever that include electrostatic \cite{sahoo08}, mechanical
by means of a dither piezo that actuates the support of the
cantilever base, magnetic \cite{elef03} and piezoelectric
\cite{pisano97}. In this article, it is assumed that the cantilever
is actuated by a dither piezo and the sensing mechanism employed is
the beam bounce method (see Figure~\ref{fig:afm}(a)).
\vspace{-.6cm}
\subsection{Models of cantilever probe, the measurement
process and the tip-media interaction}\vspace{-.2cm} A first mode
approximation of the cantilever is given by the spring mass damper
dynamics described by
\begin{equation}\label{afmdyn2}
\ddot{p}+\frac{\omega_0}{Q}\dot{p}+\omega_0^2p=\mathfrak{f}(t),\
y=p+ \upsilon,\end{equation}
 where $\ddot{p}=\frac{d^2p} {dt^2}$, $p, \mathfrak{f},\  y$ and $\upsilon$ denote the deflection of
the tip, the force on the cantilever, the measured deflection and
the measurement noise respectively whereas the parameters $\omega_0$
and $Q$ are the first modal frequency (resonant frequency) and the
quality factor of the cantilever respectively.  
The input-output transfer function with input $\mathfrak{f}$ and
output $p$ is given as
$G=\frac{1}{s^2+\frac{\omega_0}{Q}s+\omega_0^2 }.$ The cantilever
model described above can be identified precisely (see
\cite{salapakaBLMM97}). 

The interaction force, $h$, between the tip and the media depends on
the deflection $p$ of the cantilever tip. Such a dependence
 is well
 characterized by the Lennard-Jones like force that is  typically characterized by weak long-range attractive forces and strong
 short range
  repulsive forces  (see Figure~\ref{fig:afm}(c)). Thus, the probe based data storage system can
   be  viewed as an
interconnection of a linear cantilever system $G$ with the nonlinear
tip-media interaction forces in feedback  (see
Figure~\ref{fig:afm}(b) and note that $p=G(h+\eta+g)$ with
$h=\phi(p)$ \cite{sebastianSCC01}).

\vspace{-.6cm}
\subsection{Cantilever-Observer Model}
\begin{figure}[htbp]
\centering

\includegraphics[width=2.5in]{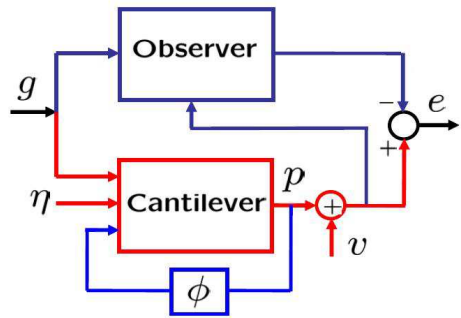}\\
 \caption{\label{fig:transDyn} An observer
architecture for the system in Figure~\ref{fig:afm}(b)}
\end{figure}

\vspace{-.2cm}
A state space representation of the filter $G$ can be obtained as $
\dot{\overline{x}}=A\overline{x}+B\mathfrak{f},\
y=C\overline{x}+\upsilon$ where $\overline{x}=[p~\dot{p}]^T$ and
$\mathfrak{f}=\eta+g$ (assuming no media forces $h$) and $A$, $B$
and $C$ are given by,
\begin{equation*}\label{ABCpara}
 A= \left[ \begin{array}{cc}
0 & 1  \\
-\omega_0^2 & -\omega_0/Q
\end{array} \right],\ \ B=\left[ \begin{array}{c}
0\\
1
\end{array} \right],\ \ C=\left[ \begin{array}{cc}
1 & 0
\end{array} \right]
\end{equation*}
  Based on the model of the cantilever, an observer to monitor the state of the cantilever
   can be implemented \cite{KailathSH} (see Figure~\ref{fig:transDyn}). The observer dynamics and the associated state estimation error
dynamics  is given by,
\[\begin{array}{l}
\overbrace{\begin{array}{lll}
\dot{\hat{\overline{x}}}&=&A\hat{\overline{x}}+Bg+L(y-\hat{y});\hat{\overline{x}}(0)=\hat{\overline{x}}_0,\\
\hat{y}&=&C\hat{\overline{x}},
\end{array}}^{Observer} \overbrace{\begin{array}{lll}
\dot{\tilde{\overline{x}}}&=&A\overline{x}+B(g+\eta)-A\hat{\overline{x}}-Bg-L(y-\hat{y}),\\
&=&(A-LC)\tilde{\overline{x}}+B\eta-L\upsilon,\\
\tilde{\overline{x}}(0)&=&\overline{x}(0)-\hat{\overline{x}}(0),
\end{array}}^{State\ Estimation\ Error\  Dynamics}

\end{array}\] where $L$ is the gain of the observer, $\hat{\overline{x}}$ is the estimate of the state
$\overline{x}$ and $g$ is the external known dither forcing applied
to the cantilever. The error in the estimate is given by
$\tilde{\overline{x}}=\overline{x}-\hat{\overline{x}}$, whereas the
error in the estimate of the output $y$ is given by,
$e=y-\hat{y}=C\tilde{\overline{x}}+\upsilon.$ The error between the
observed state and the actual state of the cantilever, when no noise
terms or media forces are present ($\eta=\upsilon=h=0$) is only due
to the mismatch in the initial conditions of the observer and the
cantilever-tip. Note that the cantilever tip interacts with the
media only for a small portion of an oscillation. It is shown in
\cite{sahooSS05} that such a tip-media interaction can be modeled
well as an impact force (in other words as an impulsive force) on
the cantilever that translates into an initial condition reset of
the cantilever state. 
The error process is white if the Kalman gain is used for $L$
\cite{KailathSH}. For cantilever deflection sensors with low enough
and realizable levels of measurement noise, the effective length of
the impulse response of the system with media force as input and the
error signal $e$ as the output can be made as short as four periods
of the cantilevers first resonant frequency.

As described in \cite{sahooSS05}, the discretized model of the
cantilever dynamics is given by
\begin{align}\label{mediaPresent}
x_{k+1} = Fx_k + G(g_k + \eta_k)+\delta_{\theta,k+1}\nu~,~
  y_k = Hx_k + v_k, \mbox{$k\geq 0$}~,
\end{align}
where the matrices $F$, $G$, and $H$ are obtained from matrices $A$,
$B$ and $C$ using the zero order hold discretization at a desired
sampling frequency and $\delta_{i,j}$ denotes the dirac delta
function. $\theta$ denotes the time instant when the impact between
the cantilever tip and the media occurs and $\nu$ signifies the
value of the impact. The impact results in an instantaneous change
or jump in the state by $\nu$ at time instant $\theta$. When a
Kalman observer is used, the profile in the error signal due to the
media can be pre-calculated as,
\begin{equation}\label{residual}
e_k = y_k - \hat{y}_k\ = \Gamma_{k;\theta}\ \nu+n_k\ ,
\end{equation}
where $\{\Gamma_{k;\theta}\ \nu\}$ is a known dynamic state profile
with an unknown arrival time $\theta$ defined by $ \Gamma_{k;\theta}
= H(F-L_KH)^{k-\theta},\text{~for~} \hspace{1mm}k \ge \theta $.
$L_K$ is the Kalman observer gain, ${n_k}$ is a zero mean white
noise sequence which is the measurement residual had the impact not
occurred and $\theta$ is assumed to be equal to 0 for simplicity.
The statistics of $n$ are given by, $ E\{n_jn_k^T\} = V\delta_{jk} $
where $V=HP_{\tilde{\overline x}}H^T+R$ and $P_{\tilde{\overline
x}}$ is the steady state error covariance obtained from the Kalman
filter that depends on $P$ and $R$ which are the variances of the
thermal noise and measurement noise respectively. 
\vspace{-.6cm}
\section{Channel model and detectors}\label{sec:channel_modelling}
\subsection{Reformulation of state space representation}\vspace{-.2cm}
It is to be noted that although we have modeled the cantilever
system as a spring-mass-damper model (second order system with no
zeros and two stable poles)(see~(\ref{afmdyn2})), the experimentally
identified channel transfer function that is more accurate in
practice has right half plane zeros that are attributed to delays
present in the electronics. Given this scenario, the state space
representation used in \cite{sahooSS05} leads to a discrete channel
with two inputs as seen in (\ref{residual}) because the structure of
$B$ is no longer in the form of $[0~ 1]^T$. However, source
information enters the channel as a single input as the tip-medium
interaction force. The problem can be reformulated as one of a
channel being driven by a single input by choosing an appropriate
state space representation. For the state space model of the
cantilever, it is known that the pair $(A,B)$ is controllable which
implies there exists a transformation which will convert the state
space into a controllable canonical form such that $B= [0~ 1]^T$.
This kind of structure of $B$ will force the discretized model
(\ref{mediaPresent}) to be such that one component of $\nu$ is equal
to $0$. With $B$ chosen as above, the entire system can be
visualized as a channel that has a single source. In this article,
the single source model is used as it simplifies the detector
structure and analysis substantially. \vspace{-.6cm}
\subsection{Channel Model}
\begin{figure}
\center
\includegraphics[height=1in,width=5in]{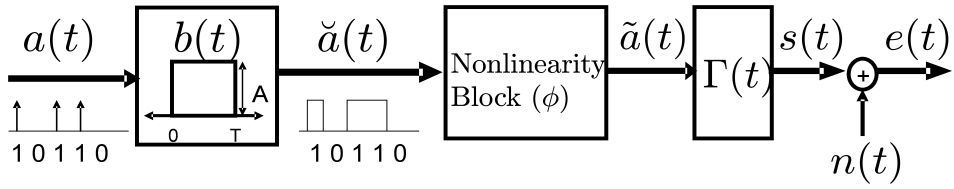}
 \caption{\label{fig:channel_model(a)} Continuous time channel model of probe based data storage
 system}
\end{figure}

\vspace{-.2cm}
The cantilever based data storage system can be modeled as a
communication channel as shown in Figure~\ref{fig:channel_model(a)}.
The components of this model are explained below in detail.
\begin{list}{}{\leftmargin=0.0cm \labelwidth=0cm \labelsep = 0cm}
\item {\bf Shaping Filter ($b(t)$)}: The model takes as input the bit sequence $\bar{a}=(a_0,~ a_1 \dots ~a_{N-1})$
where $a_{k}, k = 1, \ldots, N-1$ is equally likely to be 0 or 1. In
the probe storage context, `0' refers to the topographic profile
being {\it low} and  `1' refers to the topographic profile being
{\it high}. Each bit has a duration of $T$ seconds. This duration
can be found based on the length of the topographic profile
specifying a single bit and the speed of the scanner. The height of
the high bit is denoted by $A$. The cantilever interacts with the
media by gently tapping it when it is high. When the media is low,
typically no interaction takes place. We model the effect of the
medium height using a filter with impulse response $b(t)$ (shown in
Figure \ref{fig:channel_model(a)}) that takes as input, the input
bit impulse train $a(t) = \sum_{k=0}^{N-1} a_{k} \delta (t-kT)$. The
output of the filter is given by
$\breve{a}(t) =\sum_{k=0}^{N-1} a_{k}  b(t-kT)$.
\item {\bf Nonlinearity Block ($\phi$)}: The cantilever oscillates at frequency $f_0$
 which means that in each cantilever cycle of duration
$T_c=1/f_0$, the cantilever hits the media at most once if the media
is high during a time $T_c$. Due to the dynamics of the system
it may not hit the media, even if it is high. 
The magnitude of impact on the media is not constant and changes
according to the state of the cantilever prior to the
interaction with the media. 
We note that a very accurate modeling of the cantilever trajectory
will require the solution of complex nonlinear equations
corresponding to the cantilever dynamics and knowledge of the bit
profile so that each interaction is known. 
In this work we model the impact values of the tip-media interaction
by means of a probabilistic
Markov model that depends on the previous bits. This obviates the need for a detailed model. 

We assume that in each high bit duration $T$, the cantilever hits
the media $q$ times (i.e. $T=qT_c$) with varying magnitudes.
Therefore, for $N$
bits, 
the output of the nonlinearity block is given by, $\tilde{a}(t) =
\sum_{k=0}^{Nq-1} \nu_{k}(\bar{a}) \delta(t -k T_c)$, where
$\nu_{k}$ denotes the magnitude of the $k^{th}$ impact of the
cantilever on the medium. Here, we approximate the nonlinearity
block output as a sequence of impulsive force inputs to the
cantilever. The strength of the impulsive hit at any instant is
dependent on previous impulsive hits; precisely because the previous
interactions affect the amplitude of the oscillations that in turn
affect how hard the hit is at a particular instance. The exact
dependence is very hard to model deterministically and therefore we
chose a Markov model, as given below for the sequence of impact
magnitudes for a single bit duration,
\begin{equation}
\label{eq:non_linear_block_mem} \bar{\nu}_i=
\bar{\mathfrak{G}}(a_i,a_{i-1},\dots, a_{i-m})+\bar{\mathfrak{b}}_i
\end{equation}
\noindent where $\bar{\nu}_i=[\nu_{iq}~\nu_{iq+1} \dots
\nu_{(i+1)q-1}]^T$ and $\bar{\mathfrak{G}}(a_i,a_{i-1},\dots,
a_{i-m})$ is a function of the current and the last $m$ bits. Here
$m$ denotes the system memory and
$\bar{\mathfrak{b}}_i$ is a zero mean i.i.d. Gaussian vector of length $q$. The appropriateness of the model will be demonstrated by our experimental results. 
%
\item  {\bf Channel Response ($\Gamma(t)$)}: The Markovian modeling of the output of the nonlinearity block
as discussed above allows us to break the feedback loop in
Figure~\ref{fig:transDyn} (see also \cite{sahooSS05}). The rest of
the system can then be modeled by treating it as a linear system
with impulse response $\Gamma(t)$. $\Gamma(t)$ is the error between
the cantilever tip deflection and the tip deflection as estimated by
the observer when the cantilever tip is subjected to an impulsive
force. It can be found in closed form for a given set of parameters
of cantilever-observer system (see (\ref{residual})).
\item {\bf Channel Noise ($n(t)$)}: The measurement noise (from the imprecision in measuring the cantilever position)
and thermal noise (from modeling mismatches) can be modeled by a
single zero mean white Gaussian noise process ($n(t)$) with power
spectral density equal to $V$.
\end{list}
The continuous time innovation output $e(t)$ becomes, $e(t)
=s(t,\bar{\nu}(\bar{a})) + n(t),$ where $s(t,\bar{\nu}(\bar{a})) =
\sum_{k=0}^{Nq-1}\nu_{k}(\bar{a}) \Gamma(t-k T_c) $ and
$\bar{\nu}(\bar{a})=(\nu_0(\bar{a}),~
\nu_1(\bar{a})\dots~\nu_{Nq-1}(\bar{a}))$. The sequence of impact
values $\bar{\nu}_i$ is assumed to follow a Markovian model as
explained above, $\Gamma(t)$ is the channel impulse response and
$n(t)$ is a zero mean white Gaussian noise process.
\vspace{-.6cm}
\subsection{Sufficient Statistics for Channel model}
\begin{figure}
\center
\includegraphics[width=5in]{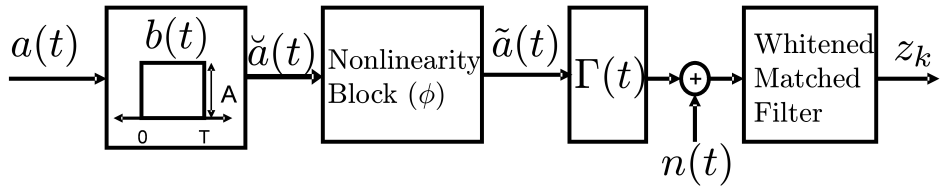}
 \caption{\label{fig:channel_model(b)} Discretized channel model with whitened matched filter
 }\vspace{-.5cm}
\end{figure}

\vspace{-.2cm}
Before providing sufficient statistics we consolidate the notation
used. The  source stream is $N$ elements long ($\bar{a}$ denotes the
sequence of source bits), with the topographic profile and the scan
speed is chosen such that the cantilever impacts any  topographic
profile $q$ times. Thus there are $Nq$ possible hits with
$\bar{\nu}(\bar{a})$ denoting the sequence of strength of the $Nq$
impulsive hits on the cantilever. Furthermore, the set of strengths
of impulsive force inputs, which is $q$ elements long,  during the
$i^{th}$ topographic profile encoding the $i^{th}$ source symbol is
denoted by $\bar{\nu}_i$. Given the probabilistic model on
$\bar{\nu}$ and finite bit sequence ($\bar{a}$), an information
lossless decomposition of $e(t)$ by expansion over an orthonormal
finite-dimensional basis with dimension $\tilde{N}$ can be achieved
where $\tilde{N}$ orthonormal basis functions span the signal space
formed by $s(t,\bar{\nu}(\bar{a}))$. The components of $e(t)$ over
$\tilde{N}$ orthonormal basis functions are given by,
$\bar{\mathfrak{e}} = \bar{s}(\bar{\nu}(\bar{a})) + \bar{n},$ where
$\bar{\mathfrak{e}} =(\mathfrak{e}_0,~
\mathfrak{e}_1\dots~\mathfrak{e}_{\tilde{N}})$,
$\bar{s}(\bar{\nu}(\bar{a})) =(s_0,~ s_1\dots~s_{\tilde{N}})$,
$\bar{n} =(n_0,~ n_1\dots~n_{\tilde{N}})$ and $\bar{n} \sim N(0, V
I_{\tilde{N}\times \tilde{N}})$ where $I_{\tilde{N}\times
\tilde{N}}$ stands for $\tilde{N}\times \tilde{N}$ identity matrix
\cite{forney1972}.
The maximum likelihood estimate of the bit sequence can be found as
$\hat{\bar{a}} = \arg \max_{\bar{a}\in \{0,1\}^N} ~ f(
\bar{\mathfrak{e}}|\bar{a} )$ where $\hat{\bar{a}} = (\hat{a}_0,
~\hat{a}_1 \dots \hat{a}_{N-1})$ is the estimated bit sequence and
$f$ denotes a pdf. The term $f( \bar{\mathfrak{e}}|\bar{a} )$ can be
further simplified as,
\begin{align*}
&f( \bar{\mathfrak{e}}|\bar{a} ) = \int_{\bar{\nu}}f(
\bar{\mathfrak{e}}|\bar{a},\bar{\nu} )f(\bar{\nu}|\bar{a})d\bar{\nu}
=\int_{\bar{\nu}} \frac{1}{{(2\pi V)}^{\frac{\tilde{N}}{2}}} \exp
[\frac{-||\bar{\mathfrak{e}} -\bar{s}(\bar{\nu}(\bar{a}))||^2 }{2V}]
f(\bar{\nu}|\bar{a})d\bar{\nu}\\
&= \frac{1}{{(2\pi V)}^{\frac{\tilde{N}}{2}}}\exp
\frac{-||\bar{\mathfrak{e}}||^2 }{2V}
 \int_{\bar{\nu}}\exp [\frac{-(||\bar{s}(\bar{\nu}(\bar{a}))||^2 - 2
\bar{\mathfrak{e}}^T \bar{s}(\bar{\nu}(\bar{a}))
)}{2V}]f(\bar{\nu}|\bar{a})d\bar{\nu}
\end{align*}
where $||.||^2$ denotes Euclidean norm, $f(
\bar{\mathfrak{e}}|\bar{a},\bar{\nu} )$ and $f(\bar{\nu}|\bar{a})$
denote the respective conditional pdf's and $\bar{\nu} = (\nu_0,~
\nu_1\dots~\nu_{Nq-1})$.
The correlation between $\bar{\mathfrak{e}}$ and
$\bar{s}(\bar{\nu}(\bar{a}))$ can be equivalently expressed as an
integral over time because of the orthogonal decomposition procedure
i.e.
$\bar{\mathfrak{e}}^T\bar{s}(\bar{\nu}(\bar{a}))=\int_{-\infty}^{\infty}
e(t) s(t,\bar{\nu}(\bar{a})) dt=\bar{\nu}^T \bar{z'}$, where
$\bar{\nu} = (\nu_0,~ \nu_1\dots~\nu_{Nq-1})$, $\bar{z'} = (z'_0,~
z'_1\dots~z'_{Nq-1})$ and $z'_{k} = \int_{-\infty}^{\infty} e(t)
\Gamma(t-kT_c) dt$ for $0 \le k \le Nq-1$ is the output of a matched
filter $\Gamma(-t)$ with input $e(t)$ sampled at $t = kT_c$. The
term $f( \bar{\mathfrak{e}}|\bar{a} )$ can now be written as,
\begin{eqnarray*}
f(\bar{\mathfrak{e}}|\bar{a} ) 
 &=& \underbrace{\frac{1}{{(2\pi
V)}^{\frac{\tilde{N}}{2}}} \exp
\frac{-||\bar{\mathfrak{e}}||^2}{2V}}
_{\mathfrak{h}(\bar{\mathfrak{e}})}
\underbrace{\int_{\bar{\nu}}\exp{
\frac{-||\bar{s}(\bar{\nu}(\bar{a}))||^2} {2V}} \exp{ \frac{
\bar{\nu}^T
\bar{z'}}{V}}f(\bar{\nu}|\bar{a})d\bar{\nu}}_{\mathfrak{F}(\bar{z'}|\bar{a})}
\end{eqnarray*}
So $f( \bar{\mathfrak{e}}|\bar{a} )$ can be factorized into
$\mathfrak{h}(\bar{\mathfrak{e}})$ (dependent only on
$\bar{\mathfrak{e}}$) and $\mathfrak{F}(\bar{z'}|\bar{a})$ (for a
given $\bar{a}$ dependent only on $\bar{z'}$). Using the
Fisher-Neyman factorization theorem \cite{stats}, we can claim that
$\bar{z'}$ is a vector of sufficient statistics for the detection
process i.e. $\frac{f(\bar{e}|\bar{a})}{f(\bar{z'}|\bar{a})} =
\mathcal{C}$, where $\mathcal{C}$ is a constant independent of
$\bar{a}$. So we can reformulate the detection problem as,
$\hat{\bar{a}} = \arg \max_{\bar{a}\in \{0,1\}^N} ~
{f(\bar{z'}|\bar{a})}$ which means that bit detection problem
depends only on the matched filter outputs ($\bar{z'}$). These
matched filter outputs for $0 \le k \le Nq-1$ can be further
simplified as, $z'_{k} = \sum_{k_1=0}^{Nq-1}\nu_{k_1}(\bar{a})
h'_{k-k_1} + n'_k$, where $h'_{k-k_1} =\int_{-\infty}^{\infty}
\Gamma(t-kT_c) \Gamma(t-k_1T_c) dt$ and $ n'_k =
\int_{-\infty}^{\infty} n(t)\Gamma(t-kT_c)dt$ such that $E(n'_k
n'_{k'}) = \int_{-\infty}^{\infty}\int_{-\infty}^{\infty}
E(n(t)n(\tau)) \Gamma(t-kT_c)\Gamma(\tau-k'T_c) dt d\tau = V
R_{k-k'}$, where $ R_{k-k'} = \int_{-\infty}^{\infty}
 \Gamma(t-kT_c)\Gamma(t-k'T_c) dt$. A whitening matched filter can be determined to whiten output noise $n'_k$ \cite{forney1972}. We shall denote the discretized output of whitened matched filter shown in Figure~\ref{fig:channel_model(b)} as $z_k$, such that
$z_k =  \sum_{k_1=0}^{I}\nu_{k-k_1}(\bar{a})h_{k_1} + n_k$, where
the filter $\{h_k\}_{k=0, 1, \dots, I}$ denotes the effect of the
whitened matched filter and the sequence $\{n_k\}$ represents the
Gaussian noise with variance $V$.
\vspace{-.6cm}
\subsection{Viterbi Detector Design}\vspace{-.2cm}
Note that the outputs of the whitened matched filter $\bar{z}$,
continue to remain sufficient statistics for the detection problem.
Therefore, we can reformulate the detection strategy as,
\begin{eqnarray}
\hat{\bar{a}} &=& \arg \max_{\bar{a}\in \{0,1\}^N} ~ f(
\bar{z}|\bar{a} )  =\arg \max_{\bar{a}\in \{0,1\}^N} ~
\Pi_{i=0}^{N-1} ~ f ( \bar{z}_i | \bar{a},\bar{z}_0^{i-1} )
\label{eq:basic_factorization}
\end{eqnarray}
where $\bar{z} = [{z_0}~{z_1} \dots {z_{Nq-1}}]^T$, $\bar{z}_i$ is
the received output vector corresponding to the  $i^{th}$ input bit,
i.e., $\bar{z_i}=[z_{iq}~z_{iq+1}\dots z_{(i+1)q-1}]^T$ and
$\bar{z}_0^{i-1}=[\bar{z}_0^T~\bar{z}_1^T \dots \bar{z}_{i-1}^T]^T$.
In our model, the channel is characterized by finite impulse
response of length $I$ i.e. $h_i = 0 ~\mbox{for $i < 0$ and $i >
I$}$ and we assume that $I \leq m_Iq$ i.e. the
inter-symbol-interference (ISI) length in terms of $q$ hits is equal
to $m_I$. Let $m$ be the system memory (see
(\ref{eq:non_linear_block_mem})). The length of channel response is
known which means that $m_I$ is known but the value of $m$ cannot be
found because it depends on the experimental parameters of the
system. In the experimental results section, we describe how we find
the value of $m$ from experimental data.
The received output vector $\bar{z}_i$ can now be written as,
\begin{align*}
& \bar{z}_i = \begin{pmatrix}
h_I & . & . & h_0 & 0 & . & . & 0 \\
0 & h_I & . & . & h_0 & 0 & . & 0  \\
\hdotsfor[4]{8}\\
 0 & . & . & 0 & h_I & . & . & h_0
\end{pmatrix} \begin{pmatrix}
\nu_{iq-I}  \\
\nu_{1+iq-I}   \\
\vdots\\
\nu_{(i+1)q-1}
\end{pmatrix} +  \bar{n}_i
= H \bar{\nu}_{i-m_I}^i + \bar{n}_i,
\end{align*}


\noindent where
$\bar{\nu}_i=[\nu_{iq}~\nu_{iq+1}\dots\nu_{(i+1)q-1}]^T$,
$\bar{\nu}_{i-m_I}^i=[\bar{\nu}_{i-m_I}^T~\dots~\bar{\nu}_{i}^T]^T$
and $ \bar{n}_i = [n_{iq}~n_{1+iq}\dots n_{(i+1)q-1}]^T$.

\begin{figure}
\center
\includegraphics[width=3.5in]{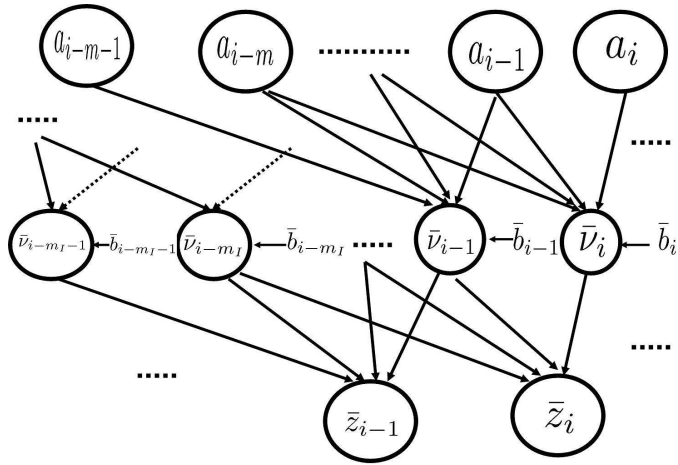}
 \caption{\label{fig:depend_graph} Dependency graph for the model with $I = m_Iq$ and $m$ is the system memory of the system}
\end{figure}

Our next task is to simplify the factorization in
(\ref{eq:basic_factorization}) so that decoding can be made
tractable.
We construct the dependency graph of the concerned quantities which
is shown in Figure~\ref{fig:depend_graph}.
Using the Bayes ball algorithm~\cite{shachter1998}, we conclude
that\vspace*{-.1mm}
\begin{align}\label{eq:factorgraph1}
&f(\bar{z}_i|\bar{\nu}_{i-m_I}^i,\bar{a},\bar{z}_0^{i-1}) =
f(\bar{z}_i|\bar{\nu}_{i-m_I}^i),\\\label{eq:factorgraph2}
&f(\bar{\nu}_{i-m_I}|\bar{a},\bar{z}_0^{i-1})=
f(\bar{\nu}_{i-m_I}|a_0^{i-1},\bar{z}_0^{i-1}),\\\label{eq:factorgraph3}
&f(\bar{\nu}_{i-k}|\bar{\nu}_{i-m_I}^{i-k-1},
\bar{a},\bar{z}_0^{i-1})=
f(\bar{\nu}_{i-k}|\bar{\nu}_{i-m_I}^{i-k-1},a_{0}^{i-m_I-1},a_{i-k-m}^{i-1},\bar{z}_0^{i-1}),\mbox{$\forall$
$1 \le k \le m_I-1$},\\& \label{eq:factorgraph4}
f(\bar{\nu}_i|\bar{\nu}^{i-1}_{i-m_I},\bar{a},\bar{z}_0^{i-1}) =
f(\bar{\nu}_i|a_{i-m}^i),
\end{align}
\noindent where $a_0^{i-1}=[a_0~a_1~\dots~a_{i-1}]$. Although the
conditional pdf
$f(\bar{\nu}_{i-k}|\bar{\nu}_{i-m_I}^{i-k-1},\bar{a},\bar{z}_0^{i-1})$
and \\$f(\bar{\nu}_{i-m_I}|\bar{a},\bar{z}_0^{i-1})$ depend on the
entire past, we assume that these dependencies are rapidly
decreasing with increase in past time. This is observed in
simulation and experimental data as well. For making the detection
process more tractable, we make the following assumptions on this
dependence,
\begin{align} \vspace{-.2cm}\label{eq:assumption1}
& f(\bar{\nu}_{i-m_I}|a_0^{i-1},\bar{z}_0^{i-1})\approx
f(\bar{\nu}_{i-m_I}|a_{i-m-m_I}^{i-1},\bar{z}_{i-m_I}^{i-1}),
\\\label{eq:assumption2}
&f(\bar{\nu}_{i-k}|\bar{\nu}_{i-m_I}^{i-k-1},a_{0}^{i-m_I-1},a_{i-k-m}^{i-1},\bar{z}_0^{i-1})
\approx
f(\bar{\nu}_{i-k}|\bar{\nu}_{i-m_I}^{i-k-1},a_{i-k-m}^{i-1},\bar{z}_{i-k}^{i-1}),
\mbox{$\forall$ $1 \le k \le m_I-1$},  \vspace{-.1cm}
\end{align}
\noindent i.e. the dependence is restricted to only the immediate
neighbors in the dependency graph. Using the above assumptions and
dependency graph results, $ f ( \bar{z_i} | \bar{a},\bar{z}_0^{i-1}
) $ can be further simplified as,\begin{align*}  &f ( \bar{z_i} |
\bar{a},\bar{z}_0^{i-1} ) = \int f ( \bar{z_i} |
\bar{\nu}_{i-m_I}^i,\bar{a},\bar{z}_0^{i-1})
 f ( \bar{\nu}_{i-m_I}^i | \bar{a},\bar{z}_0^{i-1}) d\bar{\nu}_{i-m_I}^i \\
& =\int f(\bar{z}_i|\bar{\nu}_{i-m_I}^i,\bar{a},\bar{z}_0^{i-1})
f(\bar{\nu}_{i-m_I}|\bar{a},\bar{z}_0^{i-1})
  \Pi_{k=1}^{m_I-1} f(\bar{\nu}_{i-k}|\bar{\nu}_{i-m_I}^{i-k-1},
\bar{a},\bar{z}_0^{i-1})
  f(\bar{\nu}_i|\bar{\nu}^{i-1}_{i-m_I},\bar{a},\bar{z}_0^{i-1})
  d\bar{\nu}_{i-m_I}^i\\
&= \int
f(\bar{z}_i|\bar{\nu}_{i-m_I}^i)f(\bar{\nu}_{i-m_I}|a_0^{i-1},\bar{z}_0^{i-1})
  \Pi_{k=1}^{m_I-1}f(\bar{\nu}_{i-k}|\bar{\nu}_{i-m_I}^{i-k-1},a_{0}^{i-m_I-1},a_{i-k-m}^{i-1},\bar{z}_0^{i-1})
\\ &\hspace{6cm} \times f(\bar{\nu}_i|a_{i-m}^i)d\bar{\nu}_{i-m_I}^i ~~~~\mbox{(Using (\ref{eq:factorgraph1}), (\ref{eq:factorgraph2}),(\ref{eq:factorgraph3}),(\ref{eq:factorgraph4}) )}\\
  &= \int
f(\bar{z}_i|\bar{\nu}_{i-m_I}^i)f(\bar{\nu}_{i-m_I}|a_{i-m-m_I}^{i-1},\bar{z}_{i-m_I}^{i-1})
  \Pi_{k=1}^{m_I-1} f(\bar{\nu}_{i-k}|\bar{\nu}_{i-m_I}^{i-k-1},a_{i-k-m}^{i-1},\bar{z}_{i-k}^{i-1})
 \\&  \hspace{8cm} \times f(\bar{\nu}_i|a_{i-m}^i)d\bar{\nu}_{i-m_I}^i ~~~~\mbox{(Using (\ref{eq:assumption1}),(\ref{eq:assumption2}))}\\
&=\int f ( \bar{z_i} |
\bar{\nu}_{i-m_I}^i,a_{i-m-m_I}^{i},\bar{z}_{i-m_I}^{i-1}) f (
\bar{\nu}_{i-m_I}^i | a_{i-m-m_I}^{i},\bar{z}_{i-m_I}^{i-1})
d\bar{\nu}_{i-m_I}^i
 = f ( \bar{z_i} | a_{i-m-m_I}^{i},\bar{z}_{i-m_I}^{i-1}).
\end{align*}

 \noindent By defining a state $S_i =
a_{i-m-m_I+1}^i$, this can be further expressed as $f(\bar{z}_i|S_i,
S_{i-1}, \bar{z}^{i-1}_{i-m_I})$. Again using Bayes ball algorithm,
we conclude that \vspace{-.1cm}
\begin{flalign}\label{eq:factorgraphpdf1}
&f(\bar{z}_{i-m_I}^{i}|\bar{\nu}_{i-2m_I}^i,a_{i-m-m_I}^i)
=f(\bar{z}_{i-m_I}^{i}|\bar{\nu}_{i-2m_I}^i),  &
\\\label{eq:factorgraphpdf2}
&\Pi_{k=1}^{2m_I-1}
f(\bar{\nu}_{i-2m_I+k}|\bar{\nu}_{i-2m_I}^{i-2m_I+k-1},a_{i-m-m_I}^i)
= \Pi_{k=1}^{m_I-1}
f(\bar{\nu}_{i-2m_I+k}|\bar{\nu}_{i-2m_I}^{i-2m_I+k-1},a_{i-m-m_I}^{i}) & \nn\\
&\hspace{3.5in} \times \Pi_{k=m_I}^{2m_I-1}
f(\bar{\nu}_{i-2m_I+k}|a_{i-2m_I+k-m}^{i-2m_I+k}),&
\\\label{eq:factorgraphpdf3}
&f(\bar{\nu}_{i}|\bar{\nu}_{i-2m_I}^{i-1},a_{i-m-m_I}^i) =
f(\bar{\nu}_{i}|a_{i-m}^i).&
\end{flalign}
The pdf of $\bar{z}_{i-m_I}^{i}=
[\bar{z}_{i-m_I}^T~\dots~\bar{z}_{i}^T]^T$ given current state $S_i$
and previous state
 $S_{i-1}$ is given by,
\begin{align*}
 & f(\bar{z}_{i-m_I}^{i}|S_i,S_{i-1}) =  f(\bar{z}_{i-m_I}^{i}|a_{i-m-m_I}^i)  = \int  f(\bar{z}_{i-m_I}^{i}|\bar{\nu}_{i-2m_I}^i,a_{i-m-m_I}^i)  f(\bar{\nu}_{i-2m_I}^i|a_{i-m-m_I}^i)d \bar{\nu}_{i-2m_I}^i\\
 & =\int
f(\bar{z}_{i-m_I}^{i}|\bar{\nu}_{i-2m_I}^i,a_{i-m-m_I}^i)f(\bar{\nu}_{i-2m_I}|a_{i-m-m_I}^i)
\Pi_{k=1}^{2m_I-1}
f(\bar{\nu}_{i-2m_I+k}|\bar{\nu}_{i-2m_I}^{i-2m_I+k-1},a_{i-m-m_I}^i)\\
& \times f(\bar{\nu}_{i}|\bar{\nu}_{i-2m_I}^{i-1},a_{i-m-m_I}^i) d
\bar{\nu}_{i-2m_I}^i =\int
f(\bar{z}_{i-m_I}^{i}|\bar{\nu}_{i-2m_I}^i)f(\bar{\nu}_{i-2m_I}|a_{i-m-m_I}^i)
\Pi_{k=1}^{m_I-1} f(\bar{\nu}_{i-2m_I+k}| \\ &
\bar{\nu}_{i-2m_I}^{i-2m_I+k-1},a_{i-m-m_I}^{i})
\Pi_{k=m_I}^{2m_I-1}
f(\bar{\nu}_{i-2m_I+k}|a_{i-2m_I+k-m}^{i-2m_I+k})
f(\bar{\nu}_{i}|a_{i-m}^i) d \bar{\nu}_{i-2m_I}^i \mbox{(Using
(\ref{eq:factorgraphpdf1}),(\ref{eq:factorgraphpdf2}),(\ref{eq:factorgraphpdf3}))}
\end{align*}
\noindent where the last step is obtained using results from
dependency graph and all the terms in the last step except
$f(\bar{\nu}_{i-2m_I}|a_{i-m-m_I}^i)$ and $\Pi_{k=1}^{m_I-1}
f(\bar{\nu}_{i-2m_I+k}|\bar{\nu}_{i-2m_I}^{i-2m_I+k-1},a_{i-m-m_I}^{i})$
are Gaussian distributed. 
This implies that the pdf of $\bar{z}_{i-m_I}^{i}$ given $(S_i,
S_{i-1})$ is not exactly Gaussian distributed. If the number of
states in the detector is increased it can be modeled as a Gaussian
which means that the term like $f(\bar{\nu}_{i-2m_I}|a_{i-m-m_I}^i)$
can be made Gaussian distributed by increasing the number of states,
but this increases the complexity. In order to keep the decoding
tractable we make the assumption that
$f(\bar{z}_{i-m_I}^{i}|S_i,S_{i-1})$ is Gaussian i.e.
$f(\bar{z}_{i-m_I}^{i}|S_i,S_{i-1}) \sim  N(
\bar{\mathcal{Y}}(S_i,S_{i-1}), \mathcal{C}(S_i,S_{i-1}))$, where
$\bar{\mathcal{Y}}(S_i,S_{i-1})$ is the mean and
$\mathcal{C}(S_i,S_{i-1})$ is the covariance. With our state
definition, we can reformulate the detection problem as a maximum
likelihood state sequence detection problem~\cite{kavcic2000},
\begin{align*}
 \hat{\bar{S}} &= \arg \max_{all ~\bar{S}} ~ f( \bar{z}|\bar{S} ) =
\arg \max_{all~\bar{S}} ~ \Pi_{i=0}^{N-1} ~ f (
\bar{z_i} | \bar{S},\bar{z}_0\dots \bar{z}_{i-1} ) \\
&= \arg \max_{all~\bar{S}} ~ \Pi_{i=0}^{N-1} ~ f ( \bar{z_i} | S_i,
S_{i-1},\bar{z}_{i-m_I}^{i-1} ) = \arg \max_{all~\bar{S}} ~
\Pi_{i=0}^{N-1} ~ \frac {f ( \bar{z}_{i-m_I}^{i} | S_i, S_{i-1}) }{f
(\bar{z}_{i-m_I}^{i-1} | S_i, S_{i-1})}\\
&=\arg \min_{all~\bar{S}} ~ \sum_{i=0}^{N-1} [
\log(\frac{|\mathcal{C}(S_i,S_{i-1})|}{|c(S_i,S_{i-1})|})  +
(\bar{z}_{i-m_I}^{i}-\bar{\mathcal{Y}}(S_i,S_{i-1}))^T
{\mathcal{C}(S_i,S_{i-1})}^{-1} \\
  & \times
(\bar{z}_{i-m_I}^{i}-\bar{\mathcal{Y}}(S_i,S_{i-1})) -
(\bar{z}_{i-m_I}^{i-1}-\bar{\mathbf{y}}(S_i,S_{i-1}))^T
{c(S_i,S_{i-1})}^{-1}
(\bar{z}_{i-m_I}^{i-1}-\bar{\mathbf{y}}(S_i,S_{i-1})) ]
\end{align*}
\noindent where $\hat{\bar{S}}$ is estimated state sequence,
$c(S_i,S_{i-1})$ is the upper $m_Iq\times m_Iq$ principal minor of
$\mathcal{C}(S_i,S_{i-1})$ and $\bar{\mathbf{y}}(S_i,S_{i-1})$
collects the first $m_Iq$ elements of
$\bar{\mathcal{Y}}(S_i,S_{i-1})$. It is assumed that the first state
is known. With metric given above, Viterbi decoding can be applied
to get the maximum likelihood state sequence and the corresponding bit sequence. 
\vspace{-.6cm}
\subsection{LMP, GLRT and Bayes Detector}\vspace{-.2cm}
In \cite{sahooSS05}, the hit detection algorithm is proposed which
ignores the modeling of channel memory and works well only when the
hits are sufficiently apart. In \cite{kumarciss08}, various
detectors for hit detection like locally most powerful (LMP),
generalized likelihood ratio test (GLRT) and Bayes detector are
presented. These detectors also ignore the system memory and perform
detection of single hits. Subsequently a majority type rule is used
for bit detection. The continuous time innovation ($e(t)$) is
sampled at very high sampling rate $1/T_s$ such that $T_s<<T_c$. As
the channel response ($\Gamma(t)$) is finite length, the sampled
channel response is assumed to have the finite length equal to $M$.
The sampled channel response is given by,
\begin{equation*}
\Gamma_0 = [\Gamma(t)|_{t=0}~ \Gamma(t)|_{t=T_s}\dots
~\Gamma(t)|_{t=(M-1)T_s}]^T
\end{equation*}
Determining when the cantilever is ``hitting'' the media and when it
is not, is formulated as a binary hypothesis testing problem with
the following hypotheses,
\begin{flalign*}
&H_0:\bar{e} = \bar{n},~~ H_1:\bar{e} = \Gamma_0 \nu + \bar{n}&
\label{eq:hypo_test}
\end{flalign*}
\noindent where the sampled innovation vector $\bar{e}= [e_1~e_2
\dots e_M]^T$, $\bar{n} =[n_1~n_2 \dots n_M]^T$, $\Gamma_0$ is the
sampled channel response, $\nu$ signifies the value of the impact on
media and $V I_{M\times M}$ denotes the covariance matrix of
$\bar{n}$ where $I_{M\times M}$ stands for $M\times M$ identity
matrix. In case of locally most powerful (LMP) test given in
\cite{Poor94}, the likelihood ratio is given by \cite{kumarciss08},
\begin{eqnarray*}
l_{lmp}(M) &=& \frac{\partial} {\partial \nu}( \log
\frac{f(\bar{e}|H_1)}{f(\bar{e}|H_0)})|_{\nu =0} = \bar{e}^T V^{-1}
\Gamma_0.
\end{eqnarray*}
\noindent where $l_{lmp}$ denotes likelihood ratio for LMP. In our
model, there are $q$ number of hits in one bit duration. Let
$l_{k,lmp}$ be the likelihood ratio corresponding to $k^{th}$ hit.
The decision rule for the detection of one bit in this case is
defined as,
\begin{eqnarray}\label{decisionrule}
Max\bigg{(}l_{1,lmp}(M)~,l_{2,lmp}(M)\dots
l_{q,lmp}(M)\bigg{)}\lessgtr_{1}^{0} \tau_1 \end{eqnarray}
\noindent where $\tau_1$ is LMP threshold. 
The likelihood ratio in the case of GLRT is \cite{kumarciss08},
\begin{eqnarray*}
l_{glrt}(M) &=& \log \frac{f(\bar{e}|H_1,\nu =
\tilde{\nu})}{f(\bar{e}|H_0)}= l^2_{lmp},
\end{eqnarray*}
\noindent where $\tilde \nu$ is maximum likelihood (ML) estimate of
$\nu$ i.e. $\tilde{\nu} = \arg \max_\nu f(\bar{e}|H_1)$, $l_{lmp}$
and $l_{glrt}$ are likelihood ratios for LMP and GLRT case
respectively. The decision rule for the bit detection in this case
is defined in
a similar manner given in (\ref{decisionrule}).
%

Simulations from a Simulink model of the system can be run for a
large number of hits in order to gather statistics on the
discretized output of nonlinearity block which models the tip-media
force. We modeled
the statistics for $\nu$ by a Gaussian pdf with the appropriate mean
and variance. With known mean and variance of $\nu$ the likelihood
ratio for Bayes test is \cite{kumarciss08},
\begin{eqnarray*}
l_{bayes}(M) &=& \log \frac{f(\bar{e}|H_1)}{f(\bar{e}|H_0)} =
\bar{e}^T V^{-1} \mu' + \frac {1} {2} \bar{e}^T V' \bar{e} -
\bar{e}^T V' \mu',
\end{eqnarray*}
\noindent where $\mu'=\Gamma_0 \alpha$ and $V' =\frac {\Gamma_0
\Gamma_0^T} { (\frac {V^2} {\lambda^2} + V \Gamma_0^T \Gamma_0)} $
and $\nu \sim N(\alpha, \lambda ^2)$.
The decision rule in this case is also defined in a similar manner
given in (\ref{decisionrule}). Note that $\nu$ is a measure of the
tip-medium interaction force and as such it is difficult to
experimentally verify the value of this force accurately which means
the Bayes test cannot be applied for the bit detection on actual
experimental data.
\vspace{-.6cm}
\section{Simulation Results}\label{sec:Simulation Results}\vspace{-.2cm}
We performed simulations with the following parameters. The first
resonant frequency of the cantilever
 $f_0$ = 63.15 KHz, quality factor Q =206, the value of forcing
 amplitude equal to 24 nm, tip-media separation is 28 nm, the number of hits in high bit duration is equal to $13$ i.e. $q=13$, discretized thermal and measurement noise variance are $0.1$ and $0.001$ respectively.
A Kalman observer was designed and the length of the channel impulse
response ($I$) was approximately $24$ which means that $m_I$ is
equal to $2$. We set the value of the system memory, $m=1$. Using a
higher value of $m$ results in a more complex detector. We used a
topographic profile where high and low regions denote bits `1' and
`0' respectively and the bit sequence is generated randomly.
 The simulation was performed with the above parameters using
 the Simulink model that mimics the experimental station that provides a qualitative as
well as a quantitative match to the experimental data. Tip-media
interaction was
 varied by changing the height of media corresponding to bit `1'. We define the system SNR as the nominal tip-media interaction (nm)
divided by total noise variance.

\begin{figure}
\centering
\includegraphics[width=3in]{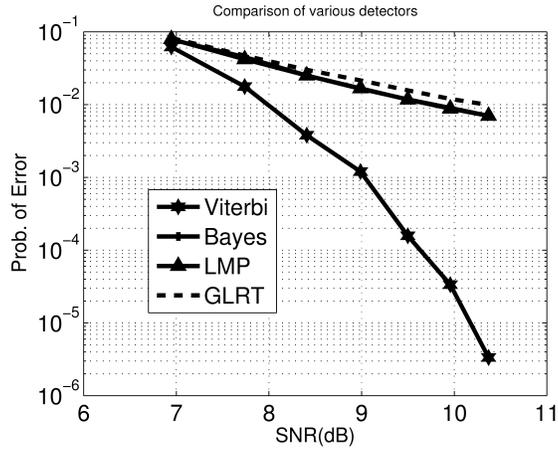}
 \caption{\label{fig:results}  Comparison of various
 detectors for simulation data. The Bayes curve is not visible in the graph as it coincides with the LMP curve.}
\end{figure}

In Figure~\ref{fig:results}, we compare the results of four
different detectors. The LMP, GLRT and Bayes detector perform hit
detection, as against bit detection. In these detectors, the system
memory is not taken into account. It is clear that the minimum
probability of error for all detectors decreases as the tip-media
interaction
 increases which makes SNR higher. The intuition behind
 this result is that hits become harder on media if tip-media interaction is
 increased which makes detection easier. The Viterbi detector gives
 best performance among all detectors because it incorporates the
 Markovian property of $\nu$ in the metric used for detection. At an SNR of 10.4 dB the Viterbi detector has a BER of $3 \times 10^{-6}$ as against the LMP detector that has $7 \times 10^{-3}$.

\vspace{-.6cm}\section{Experimental Results}\label{sec:Experimental
Results}\vspace{-.2cm} In experiments, a cantilever with resonant
frequency $f_0=71.78$~KHz and quality factor $Q=67.55$ is oscillated
near its resonant frequency. A freshly cleaved mica sheet is placed
on top of a high bandwidth piezo. This piezo can position the media
(mica sheet) in z-direction with respect to cantilever tip. A random
sequence of bits is generated through an FPGA board and applied to
the z-piezo. High level is equivalent to $1$ V and represents bit
`1' and low level is $0$ V and represents bit `0' thus creating a
pseudo media profile of $6$ nm height. The bit width can be changed
using FPGA controller from $60-350~\mu s$. The tip is engaged with
the media at a single point and its instantaneous amplitude in
response to its interaction with z piezo is monitored. The
controller gain is kept sufficiently low such that the operation is
effectively in open loop. The gain is sufficient to cancel piezo
drift and maintain a certain level of tip-media interaction. An
observer is implemented in another FPGA board which is based on the
cantilever's free air model and takes dither and deflection signals
as its input and provides innovation signal at the output. The
innovation signal is used to detect bits by comparing various bit
detection algorithms. The experiments were performed on Multimode
AFM, from Veeco Instruments. Considering a bit width of $40$ nm and
scan time of $60~\mu s$ gives a tip velocity equal to $2/3\times
10^{-3}$ m/sec. The total scan size of the media is 100 micron which
means the cantilever will take $0.15$ seconds to complete one full
scan. Read scan speed for this operation is $6.66$ Hz. The read scan
speed for different bit widths can be found in a similar manner.

The cantilever model is identified using the frequency sweep method
wherein excitation frequency $\omega$ of $g(t)=A_0\sin{\omega t}$ of
dither piezo is varied from $0-100$~KHz and $p(t)$ is recorded.
Magnitude and phase information about $G(i \omega)$ is obtained by
evaluating the ratios between steady state amplitude and phase of
output vs input excitation respectively. A second order transfer
function is obtained that best fits the experimentally identified
magnitude and phase responses of the cantilever. $A$, $B$ and $C$
matrices are obtained from the state space realization of the
identified second order transfer function. $F$, $G$ and $H$ can be
further found using the zero order hold discretization at a desired
sampling frequency. The discretized state space of the cantilever
model is used to find the discretized channel impulse response
$\Gamma_{k;\theta}$ (see (\ref{residual})).

\begin{figure}
\centering
\includegraphics[height=2.5in,width=3.5in]{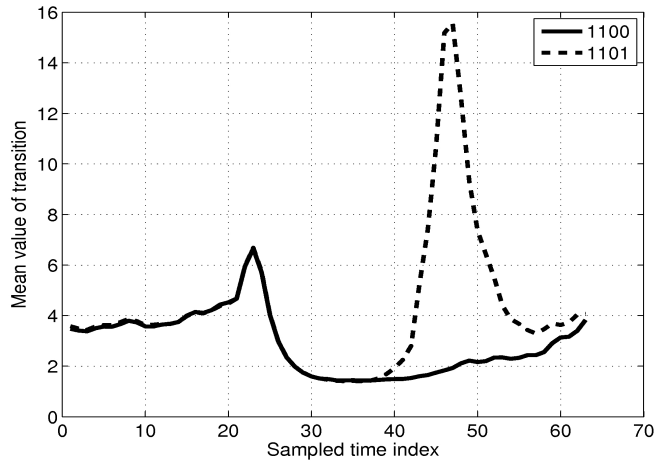}
 \caption{\label{fig:results_exp}  Mean vector for $2$ state transitions for $300~\mu s$ bit width from experimental data where `1100' and `1101' represents transition from state `110' to state `100' and `101' respectively}
\end{figure}

\begin{figure}
\centering
\includegraphics[width=3in]{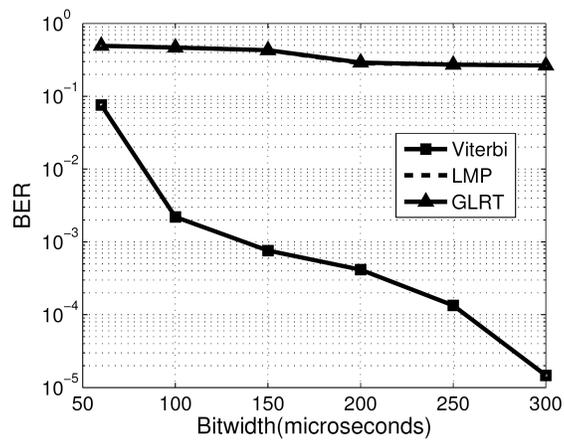}
 \caption{\label{fig:results_ber}  BER for Viterbi, LMP and GLRT for different bit widths varying from $60~\mu s$ to $300~\mu s$ for experimental data. There is a very marginal difference between LMP and GLRT curve which is not visible in the graph but LMP does perform better than GLRT.}\vspace{-.7cm}
\end{figure}

For $300~\mu s$ bit width, there are around $21$ hits in high bit
duration and Viterbi decoding is applied on the innovation signal
obtained from experiment. For experimental model, $I$ is
approximately $24$ which means $m_I$ is equal to $2$. It is hard to
estimate the system memory ($m$) from experimental parameters.
Fortunately, there is a way around for this. As shown in the
derivation of the detector, by making appropriate approximations,
the final detector only requires the mean and the covariance of each
branch in the trellis. These can be found by using training data and
assuming various values of $m$. We have varied $m$ from $0$ to $2$
and found the corresponding BER using these values of $m$. The total
number of states in the Viterbi detector is $2^{m+m_I}$. We have
observed that for $m >1$, the improvement in BER is quite marginal
as compared to the increased complexity of Viterbi decoding.
Accordingly we are using $m=1$ for which the BER from Viterbi
decoding is equal to $1\times 10^{-5}$ whereas the BER from LMP test
is $0.26$. The BER in the case of Viterbi decoding is significantly
smaller when compared to the BER for usual thresholding detectors.
If the bit width is decreased to $60~\mu s$ which means there are
around $4$ hits in the high bit duration, the BER for Viterbi
decoding is $7.56\times 10^{-2}$ whereas the BER for LMP is $0.49$
which means that LMP is doing almost no bit detection. As the bit
width is decreased, there is more ISI between adjacent bits which
increases the BER. The BER for different bit widths from all the
detectors is shown in Figure \ref{fig:results_ber}. It can be
clearly seen that Viterbi decoding gives remarkable results on
experimental data as compared to the LMP detector. The Viterbi
detector exploits the cantilever dynamics by modeling the mean and
covariance matrix for different state transitions. We have plotted
the mean vectors for $2$ state transitions with $300~\mu s$ bit
width in Figure \ref{fig:results_exp}. There are around $21$ hits in
one bit duration. The Viterbi decoding contains $8$ states and $16$
possible state transitions. In Figure \ref{fig:results_exp}, there
is a clear distinction in mean vectors for different transitions
which makes the Viterbi detector quite robust. Thresholding
detectors like LMP and GLRT perform very badly on experimental data.
For a bit sequence like `000011111', the cantilever gets enough time
to go into steady state in the beginning and hits quite hard on
media when bit `1' appears after a long sequence of `0' bits. The
likelihood ratio for LMP and GLRT rises significantly for such high
bits which can be easily detected through thresholding. However, a
sequence of continuous `1' bits keeps the cantilever in steady state
with the cantilever hitting the media mildly which means the
likelihood ratio remains small for these bits. Thus it is very
likely that long sequence of `1' bits will not get detected by
threshold detectors. \vspace{-.6cm}\section{Conclusions and future
work}\label{sec:conclusions}\vspace{-.2cm} We presented the dynamic
mode operation of a cantilever probe with a high quality factor and
demonstrated its applicability to a high-density probe storage
system. The system is modeled as a communication system by modeling
the cantilever interaction with media. The bit detection problem is
solved by posing it as a ML sequence detection followed by Viterbi
decoding. The main requirements for the proposed algorithm are (a)
the availability of training sequences which can provide the
statistics for different state transitions, (b) differences between
the tip-media interaction magnitude between `0' and `1' bit and (c)
an accurate characterization of the linear model of the cantilever
in free air. Simulation and experimental results show that the
Viterbi detector outperforms LMP, GLRT and Bayes detector and gives
remarkably low BER. The work reported in this article demonstrates
that competitive metrics can be achieved and enables probe based
high density data storage, where high quality factor probes can be
used in the dynamic mode operation. Thus, it alleviates the issues
of media and tip wear in previous probe based data storage systems.

An efficient error control coding system is a must for any data
storage system since the sector error rate specifications are on the
order of $10^{-10}$ for systems in daily use such as hard drives. In
future work, we are expecting to achieve this BER by using
appropriate coding techniques. Using run-length-limited (RLL) codes
in our system is likely to improve performance and we shall examine
this issue in future work. We are also working on a BCJR version of
the algorithm to minimize the BER of the system even further.
In experimental data, a small amount of jitter is inevitably present
which is well handled by our algorithm. At high densities, the
jitter will be significantly higher and we will need to apply more
advanced modeling and detection techniques. These are part of
ongoing and future work. \vspace{-.6cm}

\bibliographystyle{plain}

  \newpage


  \end{document}